# Tracking three-phase coexistences in binary mixtures of hard plates and spheres

Roohollah Aliabadi[1,*], Mahmood Moradi[1,†] and Szabolcs Varga[2,‡]

[1] Department of Physics, College of Science, Shiraz University, Shiraz 71454, Iran

[2] Institute of Physics and Mechatronics, University of Pannonia, PO Box 158,

Veszprém, H-8201 Hungary

Number of pages: 21 (including 4 figures and 1 table)

[*] r_aliabadi@shirazu.ac.ir and aliabadi313@gmail.com
[†] moradi@susc.ac.ir
[‡] vargasz@almos.uni-pannon.hu

**Abstract**

The stability of demixing phase transition in binary mixtures of hard plates (with thickness $L$ and diameter $D$) and hard spheres (with diameter $\sigma$) is studied by means of Parsons-Lee theory. The isotropic-isotropic demixing, which is found in mixtures of large spheres and small plates, is very likely to be preempted by crystallization. In contrast, the nematic-nematic demixing, which is obtained in mixtures of large plates and small spheres, can be stabilized at low diameter ratios ($\sigma/D$) and aspect ratios ($L/D$). At intermediate values of $\sigma/D$, where the sizes of the components are similar, neither the isotropic-isotropic nor the nematic-nematic demixing can be stabilized, but a very strong fractionation takes place between a plate rich nematic and a sphere rich isotropic phases. Our results show that the excluded volume interactions are capable alone to explain the experimental observation of the nematic-nematic demixing, but they fail for the description of isotropic-isotropic one (Chen et. al., Soft Matter, 11, 5775 (2015)).

## Introduction

Recently, there is a growing experimental and theoretical interest to understand the structure and the phase behavior of the suspension of colloidal particles. Nowadays it is possible to control the shape and the surface coverage (patchiness) of the colloidal particles and create several types of mixtures such as the rod-plate,[1,2] rod-sphere[3-5] and plate-sphere[6-11] suspensions. In these systems the anisotropic shapes are the key factors in the formation of mesophases (nematic, smectic, columnar, etc.), while the patchiness can be the driving force of gelation, glass formation and the occurrence of liquid phases at very low concentrations (empty liquid). In mixtures, both the size and the shape differences can be crucial to create materials with desired properties. In addition to this, demixing transitions can be induced with a proper choice of the shapes and sizes of the components.

The possibility of demixing phase transitions has been examined in several binary mixtures of anisotropic colloidal particles. For instance, a nematic-nematic (NN) demixing has been observed in the suspension of thick (*fd* virus coated with polyethylene glycol) and thin rods (*fd* virus), where the critical concentration of the demixing transition is above that of isotropic-nematic (IN) transition, i.e. no isotropic-nematic-nematic (INN) three-phase coexistence occurs.[12,13] Later, NN demixing transitions have also been found in plate-plate (titanite-laponite) mixture, where INN three-phase coexistence is present.[14] In rod-plate colloidal mixture, where the montmorillonite (MMT) is the plate and the sepiolite is the rod component, phase coexistences between a rod-rich nematic and a plate-rich nematic gel have been detected together with INN three-phase coexistence.[2] In the recent work of Chen et al.,[11] two types of demixing transitions have been found in the mixture of zirconium phosphate plates and silica spheres, where one is the NN, while the other is the isotropic-isotropic (II) demixing. The NN demixing takes place at the diameter ratio of $d = \sigma/D \approx 0.013$, while the II one at $d = \sigma/D \approx 0.09$, where $\sigma$ ($D$) is the diameter of the sphere (plate). It is interesting that the volume of the silica sphere is much lower that that of the plate in both cases. The hunt for demixing transitions has been started with mixing gibbsite platelets and silica spheres. In the first experiments small plates are used to induce depletion attraction between silica spheres and to produce entropic "gas-liquid" phase separation between sphere-rich and sphere-poor phases. Instead of demixing, slowing down effect of the crystallization and liquid-like microphases of the plates has been observed.[6] Later studies with small silica spheres and large gibbsite platelets have not been successful to discover either II or NN demixing transitions due to many reasons such as the gelation, columnar ordering and glass formation.[6-9] For similar reasons the binary mixture of laponite plates and silica spheres does not exhibit demixing.[15,16]

In the past, several theoretical and simulation studies have been carried out to determine the role of shape and size anisotropies in ordering and demixing transitions.[17-39] The stability of the demixing transition in plate-sphere mixture has been studied by free-volume scaled particle approach,[24] fundamental measure theory (FMT)[27,28,36] and Onsager-type theories.[36] The common point in these studies is that NN demixing may be stabilized in mixtures of small spheres and large plates. However the existence of II demixing is still under debate using different theories for mixtures of spheres and infinitely thin plates. The free volume theory, which includes the description of isotropic and solid phases only, predicts that the diameter ratio ($d$) between the spheres and infinitely thin plates must be lower than 2.44 to get demixing transition between two isotropic phases.[24] However, FMT does not confirm the existence of II demixing for any diameter ratio,[27] while the Parsons-Lee theory predicts that II can be stabilized if $d>1$.[36] The drawback of FMT and Parsons-Lee theory is that only isotropic and nematic phases have been captured with them, while the solid phase has not been considered. On the basis of the free volume theory and Parsons-Lee, one concludes that II demixing is stable for $1 < d < 2.44$ if the plate is infinitely thin. Interestingly, according to our best knowledge the effect of finite thickness of the plates on the demixing transitions has not been the subject of previous theoretical studies.

In our present study we explore the possibility of II and NN demixing transitions in binary mixtures of hard spheres and hard plates, where the aspect ratio (thickness-to-diameter ratio) of the plates is nonzero. The reason for this is that the aspect ratio of the plates in experiments exhibits large variations from 0.002 up to 0.24. By tracking of the three-phase coexistence with varying diameter ratio and aspect ratio, we can construct a global phase diagram, where the geometrical conditions for the stabilization of different demixing transitions can be visualized.

**Parsons-Lee theory**

We model the colloidal plate-sphere mixture as a binary fluid of hard spheres and hard cylinders as shown in Fig. 1. In our study the diameter of the sphere is $\sigma$, while the dimensions of the cylinder are chosen such that the length ($L$) is always lower than the diameter ($D$), i.e. the cylinder is always oblate. We examine the effects of varying aspect ratio ($k=L/D<1$) and diameter ratio ($d=\sigma/D$) on the stability of isotropic-isotropic and nematic-nematic demixing transitions using the so-called Parsons-Lee decoupling approximation for the excluded volume interactions.[40-42] From the possible three routes of the decoupling approximation for hard body mixtures[43] we employ the so-called one-fluid approximation, where the mixture is mapped into an effective hard sphere system. In this approach the free energy ($F$), which is a functional of the orientational distribution functions of the components ($f_i$, $i=p$ and $s$, where $p$ is abbreviation of the plate, while $s$ is for the sphere), can be written as

$$\beta F/V = \sum_{i=p,s} \rho_i \left(\ln \rho_i - 1 + \sigma[f_i]\right) + \frac{4-3\eta}{8(1-\eta)^2} \sum_{i,j=p,s} \rho_i \rho_j V_{exc}^{ij}[f_i, f_j], \quad (1)$$

where $\beta$ is the inverse temperature, $V$ is the volume of the system, $\rho_i = N_i/V$ is the density of component $i$, $\eta = \sum_{i=p,s}\rho_i v_i$ is the volume (or packing) fraction of the system and $v_p = \pi D^2 L/4$ ($v_s = \pi \sigma^3/6$) is the volume of a single plate (sphere). The orientational distribution functions can be found in the orientational entropy ($\sigma[f_i]$) and the excluded volume (or packing) entropy terms ($V_{exc}^{ij}[f_i, f_j]$). The first term favors the orientational disorder (isotropic phase) and is given by

$$\sigma[f_i] = \int d\omega f_i(\omega) \ln(4\pi f_i(\omega)), \qquad (2)$$

where $\omega = (\varphi, \theta)$ is a set of azimuthal and polar angles, $d\omega = \sin\theta\, d\theta\, d\phi$ and the ranges of the integrals for azimuthal and polar angles are $0 \leq \varphi < 2\pi$ and $0 \leq \theta < \pi$. The orientational distribution function ($f_i$) is normalized, i.e. $\int d\omega f_i(\omega) = 1$. Note that $4\pi$ term in Eq. (2) is just a constant, which renders $\sigma$ to be zero for isotropic distribution, where $f_i = 1/4\pi$. Since the orientational distribution of the spheres is isotropic, i.e. $f_s = 1/4\pi$, one gets that $\sigma[f_s] = 0$. However, $\sigma$ can be higher than zero for plates, i.e. $\sigma[f_p] \geq 0$, because $f_p \neq 1/4\pi$ in the nematic phase. The excluded volume term, which favors the nematic ordering, can be written as

$$V_{exc}^{ij}[f_i, f_j] = \int d\omega_1 f_i(\omega_1) \int d\omega_2 f_j(\omega_2) V_{exc}^{ij}(\vec{\omega}_1 \vec{\omega}_2), \qquad (3)$$

where $V_{exc}^{ij}(\vec{\omega}_1 \vec{\omega}_2)$ is the excluded volume between two particles of component $i$ and $j$, where the orientations of them are given by $\vec{\omega}_1$ and $\vec{\omega}_2$ orientational unit vectors. Eq. (3) can be determined analytically for sphere-sphere and sphere-plate cases, which are given by

$$V_{exc}^{ss} = 4\pi\sigma^3/3 \text{ and } V_{exc}^{ps} = V_{exc}^{sp} = \pi L(D+\sigma)^2/4 + \pi D^2\sigma/4 + \pi^2 D\sigma^2/8 + \pi\sigma^3/6. \qquad (4)$$

For plate-plate case we use the following excluded volume expression, which is derived by Onsager[44]

$$V_{exc}^{pp}(\gamma) = \left(\frac{\pi}{2}D^3 + 2L^2D\right)\sin\gamma + \frac{\pi}{2}LD^2(1+|\cos\gamma|) + 2LD^2E(\sin\gamma) \quad (5)$$

where $\gamma$ is the angle between two plates, $\cos\gamma = \vec{\omega}_1\vec{\omega}_2$, $\sin\gamma = \sqrt{1-(\vec{\omega}_1\vec{\omega}_2)^2}$ and $E(\sin\gamma) = \int_0^{\pi/2} d\phi\sqrt{1-\sin^2\gamma\sin^2\phi}$. In order to obtain the equilibrium free energy of the mixture and the orientational distribution function of the plates ($f_p$), Eq. (1) must be minimized with respect to $f_p$. We do this minimization by the iterative method of Herzfeld et al.[45]. After having obtained the free energy, we can calculate the chemical potentials ($\mu_i$) and the pressure using the standard thermodynamic relations, which are given by

$$\beta\mu_i = \frac{\partial \beta F/V}{\partial \rho_i} \text{ and } \beta P = -\beta F/V + \sum_{i=s,p}\rho_i\beta\mu_i. \quad (6)$$

Using these equations we search for isotropic-nematic, isotropic-isotropic and nematic-nematic phase separations in the mixture. Denoting with $\alpha$ and $\beta$ the coexisting phases, the coexisting component densities ($\rho_s$ and $\rho_p$ in $\alpha$ and $\beta$ phases) can be obtained from

$$\beta P^\alpha = \beta P^\beta, \quad \beta\mu_p^\alpha = \beta\mu_p^\beta \text{ and } \beta\mu_s^\alpha = \beta\mu_s^\beta \quad (7)$$

In our work we pay special attention to the aspect ratio and diameter ratio dependence of the possible three-phase coexistences to determine the stability regions of isotropic-isotropic and nematic-nematic demixing transitions. The component densities ($\rho_p^\alpha, \rho_p^\beta, \rho_p^\gamma, \rho_s^\alpha, \rho_s^\beta$ and $\rho_s^\gamma$) follows from the following six equations

$$\beta P^\alpha = \beta P^\beta = \beta P^\gamma, \quad \beta\mu_p^\alpha = \beta\mu_p^\beta = \beta\mu_p^\gamma \text{ and } \beta\mu_s^\alpha = \beta\mu_s^\beta = \beta\mu_s^\gamma. \quad (8)$$

Finally we note that the extent of nematic ordering is measured by $S_p = \int d\omega P_2(\cos\gamma)f_p(\omega)$, where $P_2(\cos\gamma) = 1.5\cos^2\gamma - 0.5$ is the second Legendre-polynomial. The $S_p$ order parameter is zero in the isotropic phase, while it reaches one in perfect nematic order (plates are parallel in this case). The order parameter of the spheres $\left(S_s = \int d\omega P_2(\cos\gamma)f_s(\omega)\right)$ is always zero.

**Results and discussion**

We have observed three different types of phase diagrams with increasing diameter ratio at $k=L/D=0.01$, which are shown in Fig. 2. We start with that phase diagram, where nematic-nematic (NN) demixing transitions occur (lower panel of Fig. 2). At $d=0.03$, one can see an isotropic-nematic (IN) phase coexistence with weak fractionation at low pressures (or volume fractions), a NN demixing transition at intermediate pressures and a reentrant isotropic-nematic transition with strong fractionation at high pressures. At these molecular parameters ($k$ and $d$), the volume of a single sphere is just 0.18% of that of a plate, i.e. adding low amount of spheres to the suspension of the plates cannot have significant impact on the IN transitions of the plates. This is the reason why the mole fraction of the spheres must exceed 0.7 or equivalently the volume fraction of them must be at least 10% of $\eta_p$ to see some destabilization effect of the spheres on IN transitions. At higher concentrations of spheres, a demixing transition emerges between two very ordered nematic phases, where one is richer in spheres than the other. This NN transition can be considered as a "vapor-liquid" transition of more or less parallel hard plates, where the presence of the spheres gives rise to effective attractive interactions between the plates, i.e. the spheres act as depletion agents. Above a threshold of sphere's concentration the NN demixing is replaced by a high density IN transitions where the isotropic phase is practically the suspension of the spheres, while the nematic phase is rich in plates, i.e. the fractionation is very strong between the coexisting two phases. This means that the high density (pressure) IN transition corresponds to a reentrant transition, because it is possible to cross two times the IN two-phase regions with increasing density at a fixed composition. At the border of IN and NN phase coexistences, the isotropic-nematic-nematic three-phase coexistence takes place. The second type of phase diagram can be observed at $d=0.2$ (middle panel of Fig. 2.), where only isotropic and nematic phases are in coexistence. The fractionation is very strong especially at high densities, where the coexisting I and N phases are practically pure in spheres and plates, respectively. The NN demixing is now metastable as the NN critical point moved down into the low density IN two-phase region. However, the reentrance phenomenon is still present, because the IN boundary can be crossed two times with increasing density at some compositions. Note that the volumes of the spheres and the plates are similar for $d=0.2$, $v_s/v_p \approx 0.53$, i.e. neither the spheres nor the plates can act as depletion agent. The very strong fractionation can be attributed to the shape incompatibility of the spheres and plates. The third type of phase diagram can be observed at higher diameter ratios, where isotropic-isotropic (II) demixing and IN transitions are present (upper panel of Fig. 2). In this case ($d=1.3$), the demixing takes place between two high-density phases, where the plates act as a depletion agent and producing attractive (depletion) interactions between the spheres. In this case the volume of the plates is just 0.68% of that of the spheres.

The stability region of the observed three types of phase diagrams can be determined by monitoring the shape and size dependence of the three-phase coexistences and the volume fractions of the phases at phase coexistences (see Fig. 3). By solving Eq. (8) at different aspect ratios and diameter ratios, one can find that $k$ and $d$ pairs where the component densities of the II or NN demixing transitions become the same. In practice we start from isotropic-isotropic-nematic (IIN) or isotropic-nematic-nematic (INN) three-phase coexistence at a given values of $k$ and $d$ and we move in $d$ upwards or downwards. We locate the stability region of demixing at that value of $d$, where the critical point of the demixing transition merges into the three-phase coexistence, i.e. two phases from the three become the same. In addition to this, the volume fractions of the coexisting three phases are monitored not to exceed either the volume fraction of hard sphere freezing for phases rich in spheres or the volume fraction of columnar ordering for phases rich in plates. In our study, the stability region of II demixing is bounded by the condition that both $\eta(I_1)$ and $\eta(I_2)$ must be lower than $\eta=0.545$, while that of NN demixing can be stable when both $\eta(N_1)$ and $\eta(N_2)$ are lower than $\eta=0.45$. Note that the hard spheres form fcc crystals for $\eta>0.545$,[46] while the hard plates get into columnar order for $\eta>0.45$.[47] The reason why we can use $\eta(N_i)<0.45$ condition is that both the isotropic-columnar and the nematic-columnar transition take place at very similar packing fractions, which are almost independent from the aspect ratio of the plates.[48] The above two rules are approximate, because the freezing and the columnar ordering are usually affected by the mixing of the components. We have performed the calculations in very wide ranges of $k$ ($0.001<k<0.1$) to cover very thin plates such as the laponite and MMT clays and thick ones. The reason why we do not consider the case of $k>0.1$, because the system of monodisperse hard plates does not form nematic phase for $k>0.1$, but it undergoes an isotropic-columnar phase transition instead of IN.[49] The regions of II and NN demixing transitions and IN phase transition are shown together with the regions of possible crystallization and columnar ordering in Fig. 3. The region of II demixing transition is not visible, because it goes together with the crystallization curve. This is due to the fact that II demixing is unlikely to occur, but it is preempted by crystallization for all molecular parameters because the volume fraction of one of the coexisting isotropic phases is always too high and located in the range of crystal phase. This shows that the excluded volume interactions are not sufficient to account for the observed isotropic-isotropic demixing transition of zirconium phosphate plate and silica sphere mixture.[11] The disappearance of the INN three-phase coexistence and the criterion for the maximum value of the packing fractions give two distinguishable curves for the stability region of NN demixing. One can see that thick plates are not suitable for NN demixing, because the region of nematic phase moves into the direction of higher densities and the columnar ordering getting competitive with increasing $k$. In order to induce NN demixing the volume of a single sphere can be even 30% of that of a single plate at $k=10^{-3}$, while it must be less than 0.4 % for $k=0.05$. However, the diameter of the spheres can not be arbitrarily low, because the NN demixing takes place at higher densities with decreasing diameter ratio. The window of suitable $\sigma$ shrinks with increasing $k$ to such an extent that NN demixing disappears completely for $k>0.08$. The molecular parameters of experimentally observed NN demixing transition[11] can be found in the region of NN demixing of our hard body model (see Fig. 3). One can also see that only IN transition is present between the crystallization (II demixing) boundary and the upper boundary of the NN demixing. Here the volumes of the components are comparable, i.e. neither the spheres nor the plates can induce demixing transitions, but the shape incompatibility is strong enough to give rise to very strong fractionations between the nematic phase rich in plates and the isotropic phase rich in spheres. At the vicinity of crystallization boundary, it may happen that the isotropic phase rich in plates is in coexistence with the solid phase of spheres. The three boundary curves of Fig. 3 can be described with 4$^{th}$ order polynomial fitting in the following form: $d_B = \sum_{i=0}^{4} A_i k^i$, where $d_B$ is the value of the $d$ at the boundary. The fitting coefficients ($A_i$) of all curves are given in Table I. Fig. 4 confirms our results for the stability regions of demixing transitions, where the volume fractions of the coexisting three phases are shown along the crystallization and the upper NN demixing boundaries of Fig. 3. Here the volume fractions of demixed phases are the same, while that of third phase is different. Regarding the II demixing (left panel of Fig. 4), one can see that the volume fractions of the isotropic phases are above 0.5, i.e. the II demixing cannot be stable with respect to freezing or glass formation. In the case of NN demixing the decreasing aspect ratio and the increasing diameter ratio are accompanied by lowering the volume fractions of the nematic-nematic critical point and that of coexisting isotropic phase along the INN boundary. Therefore the stability of NN demixing is not affected by the possible positional ordering transitions such as the columnar ordering for low values of aspect ratios and diameter ratios. However, the condition for the maximum value of the nematic packing fraction ($\eta(N_i)<0.45$) gives that the diameter ratio cannot be too low.

In summary our results are consistent with the experimental results in that sense that the binary mixture of zirconium phosphate plates and silica spheres with $d$=0.013 and $k$=0.0021, which is the only system exhibiting NN demixing transition, is located in the NN demixing region of Fig. 3, while the other experimental systems can be found in different (IN phase separation and crystallization) regions.

**Conclusions**

We have examined the role of sphere-sphere, sphere-plate and plate-plate excluded volume interactions in the stabilization of II and NN demixing phase transitions using the Parsons-Lee decoupling approximation. By tracking the IIN and INN three-phase coexistence with varying shapes and sizes for the spheres and plates, the stability regions of demixing (II and NN) transitions have been determined by means of two conditions. The first one, which prescribes that the volume fractions and mole fractions of the demixing phases to be the same, helps to determine the boundary curves of molecular systems showing INN and IIN three-phase coexistences. The second one, which does not allow the volume fractions of the coexisting phases to reach the densities of columnar and crystal phases, gives additional two boundary curves for the stability region of II and NN demixing. As the two conditions have resulted in practically identical curves for the stability region of II demixing, the probability of finding II demixing is very low in binary mixtures of hard spheres and plates. The reason for this is that the volume fractions of demixed phases are extremely high and the spheres tend to form glass or freeze. However, the polydispersity, non-uniform charge distribution and sedimentation, which are present in the experimental systems, may change the scenario by stabilizing the II demixing with respect to crystallization and other ordering. This is especially true for the recent observation of II demixing in colloidal mixtures.[10,11] Contrary to the II demixing, the theory shows that the NN demixing can be realized easily in sphere-plate mixtures as the gap between the upper and lower boundaries widens in diameter ratio with decreasing aspect ratio. The first experimental observation of NN demixing of sphere-plate mixtures[11] are in agreement with our results, since the experimental molecular parameters ($k$=0.0021 and $d$=0.013) are located in the theoretical stability region of NN demixing. The predictive power of our calculations is justified by the fact that other sphere-plate mixtures, which do not exhibit NN demixing transition, are not located in the stability region of NN demixing. It is noteworthy here that the hard-body model and the Parsons-Lee approach have already proved successful for the description of demixing transitions by accounting for the NN demixing of thick and thin *fd* viruses.[13]

To test the predictive power of our present results for II demixing, it would be useful to prepare such mixtures, where the polydispersity of both components is weak and the interactions are strongly repulsive. In this regard monodisperse platelets showing mesophases have been prepared.[50] Although the extension of the theory for more complex phases and problems such as the effect of polydispersity and the gravity is straightforward, the complexity of the density functional theory increases substantially.

**References**
[1]F. M. van der Krooij and H. N. W. Lekkerkerker, Phys. Rev. Lett. **84**, 781 (2000).
[2]P. Woolston and J. S. van Duijneveldt, Langmuir **31**, 9290 (2015).
[3]M. Adams, Z. Dogic, S. L. Keller, S. Fraden, Nature **393**, 349 (1998).
[4]G. H. Koenderink, G. A. Vliegenthart, S. G. J. M. Kluijtmans, A. van Blaaderen, A. P. Philipse and H. N. W. Lekkerker, Langmuir **15**, 4693 (1999).
[5]N. Yasarawan and J. S. van Duijneveldt, Soft Matter **6**, 353 (2010).
[6]S. M. Oversteegen, C. Vonk, J. Wijnhoven and H. N. W. Lekkerkerker, Phys. Rev. E **71**, 041406 (2005).
[7]J. C. Baird and J. Y. Walz, J. of Colloid and Interface Science **297**, 161 (2006).
[8]D. Kleshchanok, A. V. Petukhov, P. Holmqvist, D. V. Byelov, and H. N. W. Lekkerkerker, Langmuir **26**, 13614 (2010).
[9]D. Kleshchanok, J. M. Meijer, A. V. Petukhov, G. Portale, and H. N. W. Lekkerkerker, Soft Matter **7**, 2832 (2011).
[10]D. de las Heras, N. Doshi, T. Cosgrove, J. Phipps, D. I. Gittins, J. S. van Duijneveldt and M. Schmidt, Scientific Reports **2**, 789 (2012).
[11]M. Chen, H. Li, Y. Chen, A. F. Mejia, X. Wang and Z. Cheng, Soft Matter **11**, 5775 (2015).
[12]R. Purdy, S. Varga, A. Galindo, G. Jackson and S. Fraden, Phys. Rev. Lett. **94**, 057801 (2005).

[13]S. Varga, K. Purdy, A. Galindo, S. Fraden and G. Jackson, Phys. Rev. E **72**, 051704 (2005).
[14]T. Nakato, Y. Yamashita, E. Mouria and K. Kuroda, Soft Matter **10**, 3161 (2014).
[15]F. Cousin, V. Cabuil, I. Grillo and P. Levitz, Langmuir **24**, 11422 (2008).
[16]L. Bailey, H. N. W. Lekkerkerker and G. C. Maitland, Soft Matter **11**, 222 (2015).
[17]R. van Roij and B. Mulder, Phys. Rev. E **54**, 6430 (1996).
[18]R. van Roij, B. Mulder and M. Dijkstra, Physica A **261**, 374 (1998).
[19]P. C. Hemmer, Journal of Statistical Physics **100**, 3 (2000).
[20]H. H. Wensink, G. J. Vroege and H. N. W. Lekkerkerker, J. Phys. Chem. B **105**, 10610 (2001).
[21]S. Varga, A. Galindo and G. Jackson, J. Chem. Phys. **117**, 7207 (2002).
[22]D. Antypov and D. J. Cleaver, Chemical Physics Letters **377**, 311 (2003).
[23]H. H. Wensink and G. J. Vroege, J. Chem. Phys. **119**, 6868 (2003).
[24]S. M. Oversteegen and H. N. W. Lekkerkerker, J. Chem. Phys. **120**, 2470 (2004).
[25]G. Cinacchi, L. Mederos and E. Velasco, J. Chem. Phys. **121**, 3854 (2004).
[26]H. H. Wensink and G. J. Vroege, J. Phys.: Condens. Matter **16**, s2015 (2004)
[27]L. Harnau and S. Dietrich, Phys. Rev. E **71**, 11504 (2005).
[28]A. Esztermann, H. Reich, and M. Schmidt, Phys. Rev. E **73**, 011409 (2006).
[29]G. Cinacchi, Y. Martínez-Ratón, L. Mederos and E. Velasco, J. Chem. Phys. **124**, 234904 (2006).
[30]A. Cuetos, B. Martínez-Haya, S. Lago and L. F. Rull, Phys. Rev. E **75**, 061701 (2007).
[31]S. D. Peroukidis, A. G. Vanakaras and D. J. Photinos, J. Mater. Chem. **20**, 10495 (2010).
[32]J. Phillips and M. Schmidt, Phys. Rev. E **81**, 041401 (2010).
[33]N. Doshi, G. Cinacchi, J. S. van Duijneveldt, T. Cosgrove, S. W. Prescott, I. Grillo, J. Phipps and D. I. Gittins, J. Phys.: Condens. Matter **23**, 194109 (2011).
[34]G. Cinacchi, N. Doshi, S. W. Prescott, T. Cosgrove, I. Grillo, P. Lindner, J. S. Phipps, D. Gittins, and J. S. Van Duijneveldt, J. Chem. Phys. **137**, 204909 (2012).
[35]R. Berardi and C. Zannoni, Soft Matter **8**, 2017 (2012).
[36]D. de las Heras and M. Schmidt, Phil. Trans. R. Soc. A **371**, 20120259 (2013).
[37]F. Gámez, R. D. Acemel and A. Cuetos, Mol. Phys. **111**, 3136 (2013).
[38]H. Lekkerkerker, R. Tuinier and H. Wensink, Mol. Phys. **113**, 2666 (2015).
[39]L. Wu, A. Malijevsky, G. Jackson, E. A. Muller and C. Avendano, J. Chem. Phys. **143**, 044906 (2015).
[40]J. D. Parsons, Phys. Rev. A **19**, 1225 (1979).
[41]S. D. Lee, J. Chem. Phys. **87**, 4972 (1987).
[42]P. Padilla and E. Velasco, J. Chem. Phys. **106**, 10299 (1997).
[43]A. Malijevský, G. Jackson and S. Varga, J. Chem. Phys. **129**, 144504 (2008).
[44]L. Onsager, Ann. N.Y. Acad. Sci. **51**, 627 (1949).
[45]J. Herzfield, A. E. Berger, and J. W. Wingate, Macromolecules **17**, 1718 (1984).
[46]W. G. Hoover and F. H. Ree, J. Chem. Phys. **49**, 3609 (1968).
[47]S. D. Zhang, P. A. Reynolds and J. S. van Duijneveldt, J. Chem. Phys. **117**, 9947 (2002).
[48]H. H. Wensink and H. N. W. Lekkerkerker, Mol. Phys. **107**, 2111 (2009).
[49]J. A. C. Veerman and D. Frenkel, Phys. Rev. A **45**, 5632 (1992).
[50]A. B. D. Brown, S. M. Clarke, and A. R. Rennie, Langmuir **14**, 3129 (1998).

**Figures**

**FIG. 1.** Schematic representation of the components of the binary mixture: hard plate (left) and hard sphere (right). $L$ and $D$ are the thickness and the diameter of the plate, respectively. $\sigma$ is the diameter of the sphere.

**FIG. 2.** Phase diagram of the binary mixture of spheres and plates in density-density ($\eta_s - \eta_p$) and pressure-composition ($P^*$-$x$) planes: $\sigma/D$ =1.3 (upper panels, (a) and (b)), $\sigma/D$ =0.2 (middle panels, (c) and (d)) and $\sigma/D$ =0.03 (lower panels, (e) and (f)). The aspect ratio of the plates ($k=L/D$) is 0.01 in all cases. The labels I and N denote the isotropic and nematic phases, respectively. The coexisting isotropic (nematic) phases are labeled as $I_1$ and $I_2$ ($N_1$ and $N_2$). The two-phase and three-phase regions are indicated by gray (I-N), green ($I_1$-$I_2$), cyan ($N_1$-$N_2$) and pink (I-$N_1$-$N_2$ and $I_1$-$I_2$-N). The mole fraction is that of the hard spheres ($x = x_s$), while $P^* = \beta P v_p$. Dashed lines are tie lines connecting the coexisting phases.

**FIG. 3.** Stability regions of different kind of phase diagrams of sphere-plate mixtures are shown in diameter ratio-aspect ratio ($\sigma/D$- $L/D$) plane. The regions of the phase diagrams are coloured differently: 1) The phase diagram is dominated by strong fractionation and reentrance of isotropic-nematic phase transition (gray), 2) the region of isotropic-nematic and isotropic-crystal transition, where the isotropic-isotropic demixing is not stable (green), 3) the region of nematic-nematic demixing and isotropic-nematic transition (cyan) and 4) those systems where the nematic-nematic demixing is unlikely to occur and replaced by isotropic-columnar transition (white). The molecular parameters of some experimental systems are highlighted by triangle symbols and the numbers refer to their references.

**FIG. 4.** Coexisting packing fractions ($\eta$) vs. diameter ratio ($\sigma/D$) at the disappearance of the isotropic-isotropic-nematic (left panel, (a)) and isotropic-nematic-nematic three-phase (right panel, (b)) coexistences. The arrows indicate the direction of increasing aspect ratio of the plates.

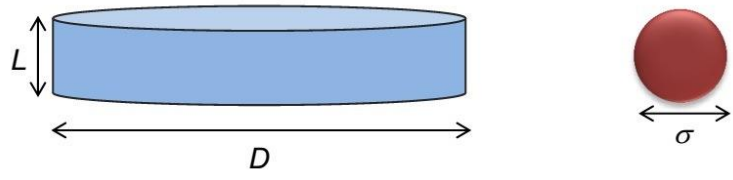

**FIG. 1**

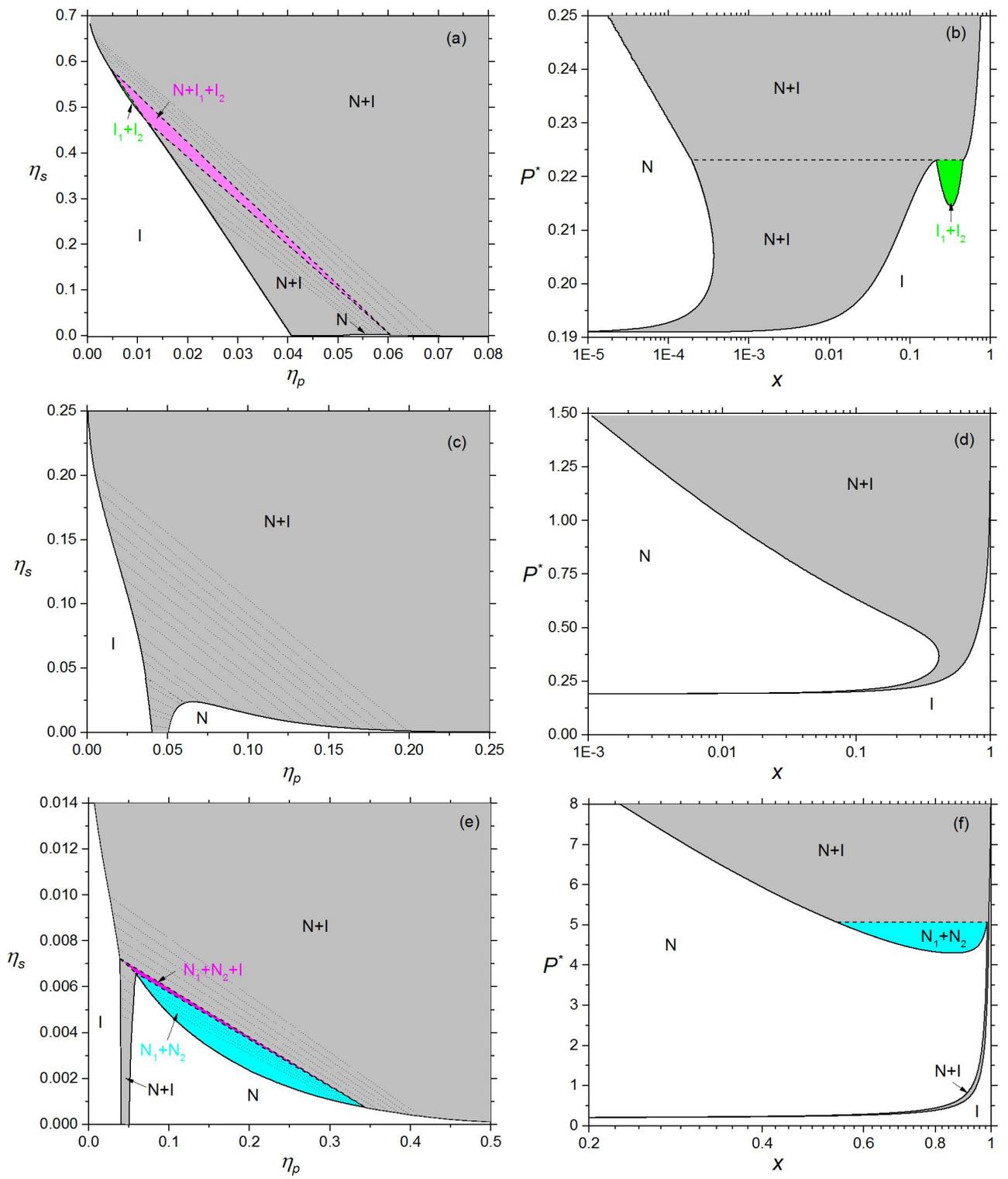

**FIG. 2**

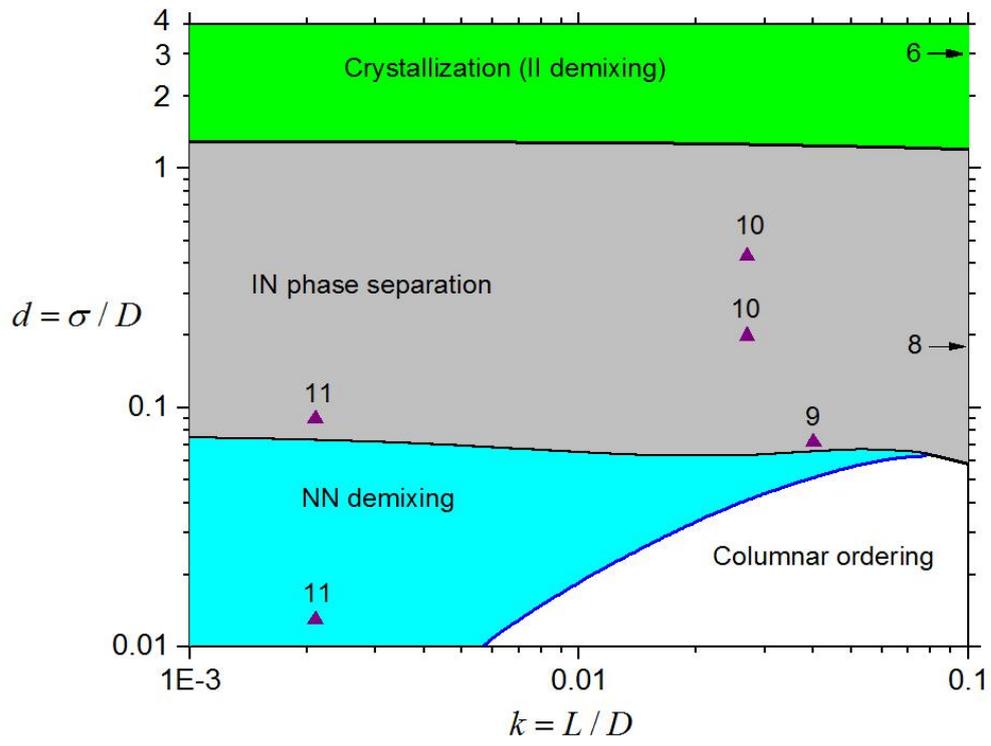

**FIG. 3**

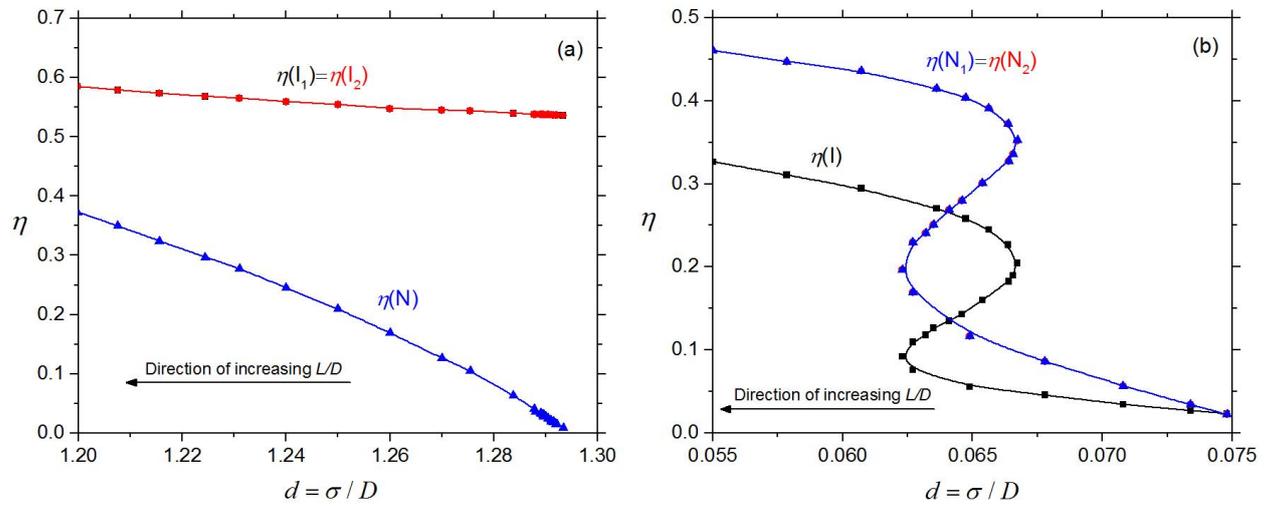

**FIG. 4**

**Table I**

Fitting coefficients of the crystallization boundary (lower boundary of II demixing), upper boundary of NN demixing and boundary of columnar order (lower boundary of NN demixing)

|  | $A_0$ | $A_1$ | $A_2$ | $A_3$ | $A_4$ | R-square |
|---|---|---|---|---|---|---|
| Crystallization boundary | 1.293765 | -0.922837 | -3.049368 | 36.740169 | -81.166645 | 0.999 |
| Upper boundary of NN demixing | 0.075841 | -1.464574 | 52.400281 | -672.680876 | 2775.198609 | 0.984 |
| Boundary of columnar order | -0.003427 | 2.673333 | -50.050256 | 533.955240 | -2466.223212 | 0.999 |